# Short Duration Traffic Flow Prediction Using Kalman Filtering

Khondhaker Al Momin[1, a)], Saurav Barua[2], Md. Shahreer Jamil[3], Omar Faruqe Hamim[4]

[1,2] Department of Civil Engineering, Daffodil International University, Bangladesh.
[3,4] Department of Civil Engineering, Bangladesh University of Engineering & Technology, Bangladesh

[a)] Corresponding Author: momin.ce@diu.edu.bd

**Abstract:** The research examined predicting short-duration traffic flow counts with the Kalman filtering technique (KFT), a computational filtering method. Short-term traffic prediction is an important tool for operation in traffic management and transportation system. The short-term traffic flow value results can be used for travel time estimation by route guidance and advanced traveler information systems. Though the KFT has been tested for homogeneous traffic, its efficiency in heterogeneous traffic has yet to be investigated. The research was conducted on Mirpur Road in Dhaka, near the Sobhanbagh Mosque. The stream contains a heterogeneous mix of traffic, which implies uncertainty in prediction. The propositioned method is executed in Python using the pykalman library. The library is mostly used in advanced database modeling in the KFT framework, which addresses uncertainty. The data was derived from a three-hour traffic count of the vehicle. According to the Geometric Design Standards Manual published by Roads and Highways Division (RHD), Bangladesh in 2005, the heterogeneous traffic flow value was translated into an equivalent passenger car unit (PCU). The PCU obtained from five-minute aggregation was then utilized as the suggested model's dataset. The propositioned model has a mean absolute percent error (MAPE) of 14.62, indicating that the KFT model can forecast reasonably well. The root mean square percent error (RMSPE) shows an 18.73% accuracy which is less than 25%; hence the model is acceptable. The developed model has an $R^2$ value of 0.879, indicating that it can explain 87.9 percent of the variability in the dataset. If the data were collected over a more extended period of time, the $R^2$ value could be closer to 1.0. The proposed estimation technique can be dynamically implemented in a travel time forecast and traffic flow value forecasting application tool. KFT has been thoroughly examined in both motorized and non-motorized vehicles. The research might be expanded to other geographical areas. Furthermore, traffic volume at a different level of service may be studied to validate this study on a broad scale.

## INTRODUCTION

Traffic flow prediction is considered as a challenging problem in Intelligent Transportation Systems, and short-term traffic foretelling has been a substantial research topic in traffic engineering since the nineteen-eighties (Abadi, Rajabioun and Ioannou, 2015). The phrase "short-term" refers to traffic forecasting for the immediate future based on historical traffic data collected over a short period of time. Short-term information improves traffic management in order to reduce congestion and provide dynamic bandwidth allocation (Rajabzadeh, Rezaie and Amindavar, 2017). The demand in a transportation system is traffic flow, and it affects vehicle speed, travel time, and the degree of service on a route. Increased traffic volume causes traffic congestion, long lines, and delays, all of which negatively impact the pavement (Momin and Hamim, 2022). Short-term traffic flow forecast may be integrated with the Advanced Traveller Information System, which offers route guiding services, multi-modal trip planning, and advising functions to passengers. Furthermore, predicting short-term traffic flow helps estimate travel time and the existing level of service on a route.

Traffic prediction models can be divided into three groups based on their state-of-the-art (Van Hinsbergen, Van Lint and Sanders, 2007). The first type is comprised of Naïve methods, in which the models do not make use of any unique intelligence in order to anticipate the target values. The historical average is a good illustration of one of these approaches. The performance of these approaches in terms of forecasts, on the other hand, is relatively poor. The second type of approach is known as parametric methods, and it is comprised of models that use traffic flow theory to antedate traffic status. Finally, nonparametric approaches are an additional class of traffic forecast techniques that employ an adaptable mathematical approximation function with configurable factors. These approaches have the capacity to learn the structure and parameters of a nonlinear model from the given data. For instance, (Nicholson and Swann, 1974; Williams and Hoel, 2003; Kumar and Vanajakshi, 2015) are a few examples of this nonparametric approach.

Regression, neural networks, historical average algorithms, and time series analysis are some of the mathematical strategies that have been used for predicting traffic volume in the past (Kumar, Parida and Katiyar, 2015; Cheng et al., 2017; Rajabzadeh, Rezaie and Amindavar, 2017). For traffic flow prediction, the use of Autoregressive Integrated Moving average (ARIMA) or seasonal ARIMA (SARIMA) for traffic flow prediction models requires a huge amount of flow data for model development, and hence it may not be practical to utilize ARIMA in situations where adequate data is not available (Kumar, 2017). These limits compel the development of a different approach that might possibly be applied to the problem of traffic flow prediction while also overcoming the limitations of present traffic flow prediction models. Because of this, a Kalman filtering technique (KFT)-based prediction system was devised and assessed in this study.

This research aims to anticipate short-term traffic flow value for the urban midblock road section, and we employed KFT to do so. KFT is an algorithm that predicts certain unknown variables based on measurements taken over time. Several pieces of research related to our study have been conducted. We can find a comprehensive overview and challenges of KFT in (Vlahogianni, Karlaftis and Golias, 2014; Nadi et al., 2021). (Kumar, 2017) forecasting traffic flow count using KFT, and showed that it outperforms time series models, and studies conducted by (Yin et al., 2002; Castro-Neto et al., 2009; Abadi, Rajabioun and Ioannou, 2015; Tan et al., 2016) showed that short-term traffic flow forecast errors varied approximately 2% for a 5 minute time period

## METHODOLOGY

Developing a datasheet for traffic counting before undertaking a field survey is necessary. We chose a midblock metropolitan road section and used a video camera to capture the traffic flowing across it. Each five-minute traffic flow was aggregated, and heterogeneous traffic was converted into Passenger Car Equivalent (PCE) according to standard code. Then, the traffic counting sheet data was imported into an excel spreadsheet. The Kalman.py package from API's python library is used to forecast the traffic flow. A quantitative and graphical comparison has been conducted to determine the KFT's suitability.

### Study Area

One of Dhaka's busiest arterials, Mirpur Road, was the research site, which was close to the Sobhanbagh Mosque. The video was shot from a bridge near the Sobhanbagh Mosque as the narrator walked. Southbound traffic (traffic to Kalabagan intersection) was only counted, shown in figure 1.

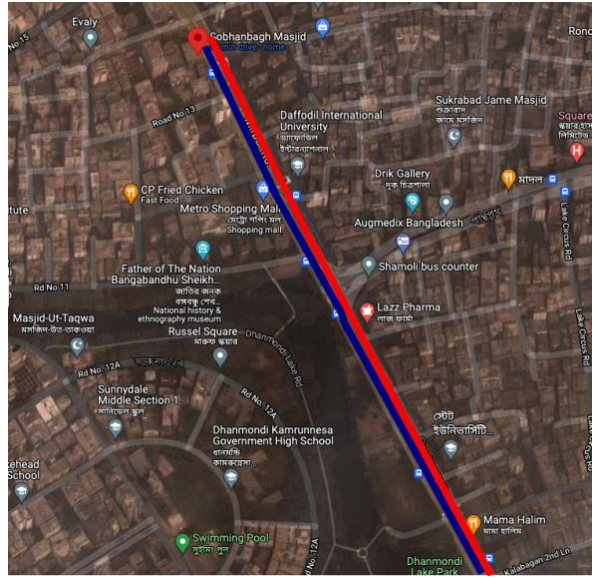

*Figure 1.* Data Collection Location on Google Map

## DATA COLLECTION

To develop and test our model, we performed a classified vehicle count on the Dhanmondi-Mirpur connecting road, one of the busiest roads in Dhaka city, on January 20, 2021, for three hours. The traffic count is carried out by manually counting the vehicles from recorded footage. During vehicle counting, traffic data is collected every 5 minutes and aggregated. Dhaka is becoming a city of Rickshaws, an NMV (Non-Motorized Vehicle) with a slower running speed that clings to traffic congestion on the city's roads (Momin, Islam and Barua, 2021); thus, it is necessary to convert different kinds of vehicles to the Equivalent Passenger Car-unit (PCU) according to the Geometric Design Standards Manual published by Roads and Highways Division (RHD), Bangladesh in 2005, shown in table 1 (Roads and Highways Division, 2005).

**TABLE 1:** PCU Converting Guideline for Different Types of Vehicles.

| Vehicle Type | PCU |
| --- | --- |
| Bus | 3.0 |
| Truck | 3.0 |
| CNG | 0.75 |
| Private Car | 1.0 |
| Commercial Vehicle | 1.0 |
| Utility | 1.0 |
| Motorcycle | 0.75 |
| Bicycle | 0.5 |
| Cycle Rickshaw | 2.0 |

### Kalman Filtering Technique

R. E. Kalman developed the Kalman Filtering Technique (KFT) in 1960 (Sorenson, 1966), which is also known as Linear Quadratic Estimation (LQE). In KFT, a sequence of measurements is observed over time, including noise (error) and added inaccuracies, and it estimates unknown variables that are more accurate than those based on one single observation. A joint probability is computed for each time step across all the variables throughout the calculation.

Assume,
$\Delta t$ = Time difference,
$F_t$ = Response variable at $\Delta t$
$M_t$ = Transition matrix
$\omega_t$ = Pragmatic noise

KFT model: $F_{t+1} = F_t \times M_t + \omega_t$ (1)

$\epsilon_t$ = Predicted variable at $\Delta t$
$M_m$ = Measurement matrix
$M_n$ = Measurement noise

Measurement model: $\epsilon_t = M_m \times F_t + M_n$ (2)

The measurement matrix is used to estimate current measurement by multiplying it with the predicted state. The measurement noise is a matrix used to adjust current data by considering previous data. It represents the data collection system's noise characteristics. The modification matrix is used to adjust the input values to estimate future values.

Projected subsequent observation = $\hat{x}^+_{\Delta t+1}$
Projected preceding observation = $\hat{x}^-_{\Delta t+1}$

KFT gain at $\Delta t+1$ time = $k_{t+1}$, where $k_{t+1}$ is determined by the combined likelihood of measurement and estimated observation.

Amended model: $\hat{x}_{\Delta t+1} = \hat{x}^-_{\Delta t+1} + k_{t+1} \times [\epsilon_t +1 - \hat{x}^-\Delta t +1]$ (3)

If $\omega_t > M_n$, the outcome variable is significantly dependent on the model, and if $\omega_t < M_n$, the outcome variable is more dependent on measurement.

## DATA ANALYSIS

The vehicle counts obtained during the five-minute period were translated into PCU values. Every five minutes, PCU data was imported into an excel spreadsheet, which was then used in Python IDE. These were the datasets that were observed (Field). The traffic flow count was modeled using a Kalman filtering module called "PyKalman," which is built-in Python version 3.8. While developing in the Python environment, additional supporting packages such as Pylab, NumPy, Matplotlib, and Seaborn were imported to aid in the development process.

### Descriptive Statistics

The average flow count in the observed and projected datasets is 488.33and 484.5, respectively, in the observed dataset. The standard deviation of the observed dataset is 180.68, whereas the standard deviation of the projected dataset is 176.35. The observed dataset has a slightly larger flow count than the anticipated dataset, which is consistent with previous findings. The histograms of observed and predicted traffic flow values are shown in figure 2 and 3, respectively.

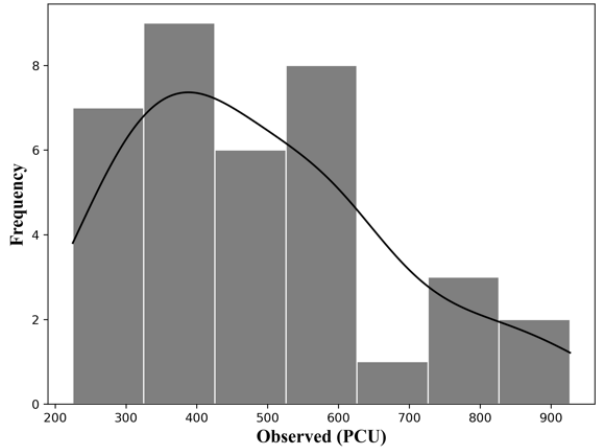 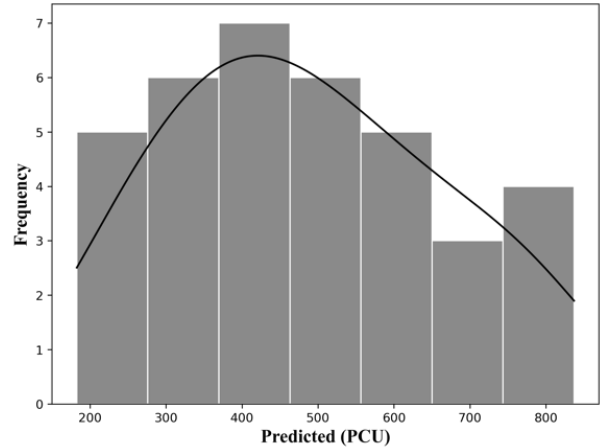

*Figure 2:* Histogram of Observed Traffic Flow     *Figure 3:* Histogram of Predicted Traffic Flow

It is found that the observed dataset of flow count has a range of 225 – 927 PCU, whereas the projected dataset of flow count has a range of 182.62 – 837 PCU, respectively. When compared to the actual dataset, the observed dataset has a more extensive range. Figure 4 displays a boxplot of the observed and predicted flow values, where the median value 465.72 PCU for both observed and predicted are equal, indicating the efficacy of the model.

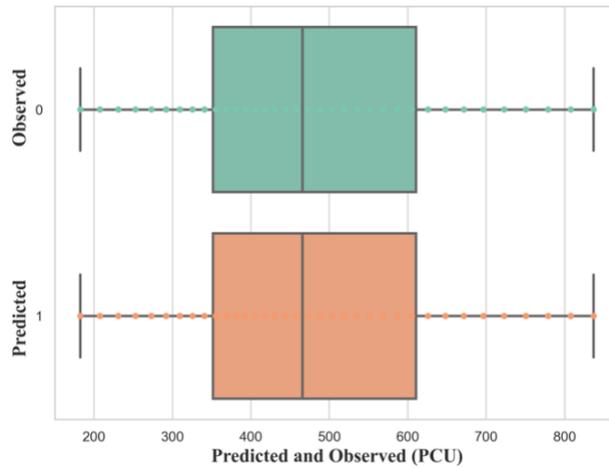

*Figure 4:* Histogram of Predicted Traffic Flow

**Model Comparison**

Since most of the points in the graph, predicted traffic flow versus observed traffic flow, shown in figure 5, are incredibly close to the line passing through the origin, it justifies the efficacy of the model with a person correlation value of 0.937 and $R^2$ value of 0.879, indicating that it can explain 87.9 percent of the variability in the dataset. In figure 6, the KFT analysis reveals that the observed traffic flow values have a growing trend, indicating the road section is becoming farther crowded over time, as seen during spot observation.

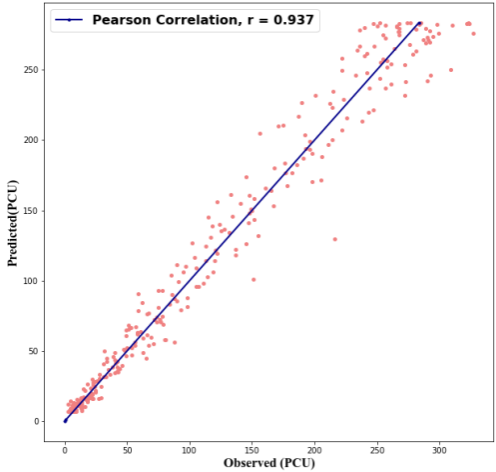
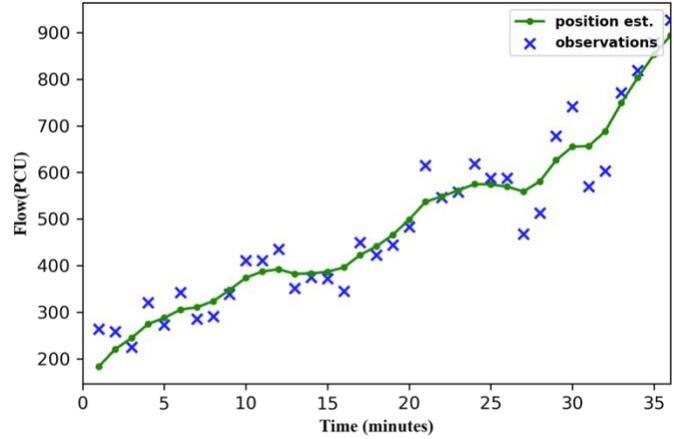

*Figure 5:* *Pearson correlation value, observed vs. predicted value*

*Figure 6:* *Trend forecasting by KFT model*

**Mean Absolute Percent Error (MAPE) and Root Mean Square Percent Error (RMSPE)**

The Mean Absolute Percentage Error (MAPE) measures how accurate a prediction system is. MAPE and Root Mean Square Percent Error (RMSPE) are two metrics considered to estimate the efficacy of a forecasting technique (Yang, Liu and You, 2010; Wang *et al.*, 2014; De Myttenaere *et al.*, 2016).

$$\text{MAPE} = \frac{100}{n} \sum_{i=1}^{n} \left| \frac{x_i' - x_i}{x_i'} \right| \qquad (4)$$

$$\text{RMSPE} = \sqrt{\frac{\sum_{i=1}^{n}(\frac{x_i' - x_i}{x_i'})^2}{n}} \times 100 \qquad (5)$$

Where, $x_i'$ = Forecasted traffic flow value,
$x_i$ = Input Traffic flow value, and
n = sample size

In our model, the value of MAPE is 14.62%, and RMSPE is 18.73%. High-accuracy forecasting is defined as MAPE 10 percent; 10-20 percent is regarded as excellent forecasting; 20-50 percent is considered decent forecasting; and >50 percent is deemed to be bad predicting (Tayman and Swanson, 1999; Prayudani *et al.*, 2019). RMSPE 25 percent indicates that the model is acceptable, >25 percent shows that re-calibration is required, and 100 percent indicates that the model and observed data are in extremely poor agreement (Wang *et al.*, 2014; De Myttenaere *et al.*, 2016). The standard values of MAPE and RMSPE provided a clear indication of the degree of precision of our model.

## RESULT AND DISCUSSION

The standard deviation of the observed dataset is 180.68, while the standard deviation of the predicted dataset is 176.35. The $R^2$ of the KFT predicted flow count is determined to be 0.879 while equated to the field traffic flow count. The gradient of the KFT-forecasted dataset is fairly positive, rising with the number of observations. The model has a MAPE value of 14.62% and an RMSPE value of 18.73%, demonstrating that the suggested model is acceptable. The predicted dataset has less standard deviation, variance, and range than the pragmatic dataset. The model predicts that the study road portion will become more congested over the time. The MAPE indicates that KFT has excellent forecasting ability to estimate short-term traffic flow value, and the RMSPE test demonstrates that the proposed model is acceptable. For noise and disturbance measurements in the dataset, KFT employs a Gaussian distribution. The use of advanced filtering techniques, such as extended Kalman filtering, unscented filtering, Stochastic filtering, Bayesian filtering, and particle filtering in conjunction with Kalman filtering, might help increase the model's accuracy. The suggested KFT may improve the accuracy of route recommendations by incorporating it into the software in real-time.

## CONCLUSIONS

The goal of this research is to forecast a five-minute aggregated traffic count over a three-hour period using the KFT. The count of short-term traffic flow is critical in Intelligent Transportation Systems. It directs the traveler's path and dynamically gives travel directions to users. On the other hand, long-term traffic flow varies seasonally, so for dynamic traffic operation, a short duration traffic flow count is necessary.

KFT traffic flow prediction may be used to compare different routes. It may help determine the nature of traffic blockage with its queue length. The Global Positioning System, loop detectors, and other measuring methods might be readily combined with video footage data to provide more precise traffic flow predictions. The suggested KFT model can be evaluated for long-duration traffic flow estimation and compared to time series models for heterogeneous road traffic systems.

## ACKNOWLEDGEMENT

The authors of this paper would like to thank the Department of Civil Engineering, Daffodil International University, for the technical assistance in conducting this research.

## REFERENCES

1. Abadi, A., Rajabioun, T. and Ioannou, P. A. "Traffic Flow Prediction for Road Transportation Networks With Limited Traffic Data," *IEEE Transactions on Intelligent Transportation Systems*, 16(2), pp. 653–662 (2015). doi: 10.1109/TITS.2014.2337238.
2. Castro-Neto, M. *et al.* "Online-SVR for short-term traffic flow prediction under typical and atypical traffic conditions," *Expert Systems with Applications*, 36(3 PART 2), pp. 6164–6173 (2009) doi: 10.1016/j.eswa.2008.07.069.
3. Cheng, A. *et al.* "Multiple sources and multiple measures based traffic flow prediction using the chaos theory and support vector regression method," *Physica A: Statistical Mechanics and its Applications*, 466, pp. 422–434 (2017), doi: 10.1016/j.physa.2016.09.041.
4. Van Hinsbergen, C. P., Van Lint, J. W. and Sanders, F. M. "Short term traffic prediction models," in *Proceedings Of The 14th World Congress On Intelligent Transport Systems (Its), Held Beijing, October (2007)*.
5. Kumar, K., Parida, M. and Katiyar, V. K. "Short term traffic flow prediction in heterogeneous condition using artificial neural network," *Transport*, 30(4), pp. 397–405 (2015). doi: 10.3846/16484142.2013.818057.
6. Kumar, S. V. "Traffic Flow Prediction using Kalman Filtering Technique," *Procedia Engineering*, 187, pp. 582–587 (2017). doi: 10.1016/j.proeng.2017.04.417.
7. Kumar, S. V. and Vanajakshi, L. "Short-term traffic flow prediction using seasonal ARIMA model with limited input data," *European Transport Research Review*, 7(3), pp. 1–9 (2015). doi: 10.1007/s12544-015-0170-8.
8. Momin, K. A. and Hamim, O. F. "Pavement Management System Using Deflection Prediction Model of Flexible Pavements in Bangladesh," in *Lecture Notes in Civil Engineering*, pp. 363–370 (2022). doi: 10.1007/978-981-16-5547-0_34.
9. Momin, K. A., Islam, S. and Barua, S. "Effects of Non-motorized Vehicle on Speed–Density Relationship of An Urban Intersection," *Trends in Transport Engineering and Applications*, 8(1), pp. 22–29 (2021).
10. De Myttenaere, A. *et al.* "Mean Absolute Percentage Error for regression models," *Neurocomputing*, 192, pp. 38–48 (2016). doi: 10.1016/j.neucom.2015.12.114.
11. Nadi, A. *et al.* "Short-term prediction of outbound truck traffic from the exchange of information in logistics hubs: A case study for the port of Rotterdam," *Transportation Research Part C: Emerging Technologies*, 127(March), pp. 1–18 (2021). doi: 10.1016/j.trc.2021.103111.

12. Nicholson, H. and Swann, C. D. "The prediction of traffic flow volumes based on spectral analysis," *Transportation Research*, 8(6), pp. 533–538 (1974). doi: 10.1016/0041-1647(74)90030-6.
13. Prayudani, S. *et al.* "Analysis Accuracy of Forecasting Measurement Technique on Random K-Nearest Neighbor (RKNN) Using MAPE and MSE," *Journal of Physics: Conference Series*, 1361(1) (2019). doi: 10.1088/1742-6596/1361/1/012089.
14. Rajabzadeh, Y., Rezaie, A. H. and Amindavar, H. "Short-term traffic flow prediction using time-varying Vasicek model," *Transportation Research Part C: Emerging Technologies*, 74, pp. 168–181 (2017). doi: 10.1016/j.trc.2016.11.001.
15. Roads and Highways Division. "*Geometric Design Standards Manual (Revised)*". Dhaka: Roads and Highways Division (RHD) (2005).
16. Sorenson, H. W. "Kalman Filtering Techniques," in. Elsevier, pp. 219–292 (1966). doi: 10.1016/b978-1-4831-6716-9.50010-2.
17. Tan, H. *et al.* "Short-Term Traffic Prediction Based on Dynamic Tensor Completion," *IEEE Transactions on Intelligent Transportation Systems*, 17(8), pp. 2123–2133 (2016). doi: 10.1109/TITS.2015.2513411.
18. Tayman, J. and Swanson, D. A. "On the validity of MAPE as a measure of population forecast accuracy," *Population Research and Policy Review*, 18(4), pp. 299–322 (1999). doi: 10.1023/A:1006166418051.
19. Vlahogianni, E. I., Karlaftis, M. G. and Golias, J. C. "Short-term traffic forecasting: Where we are and where we're going," *Transportation Research Part C: Emerging Technologies*, 43, pp. 3–19 (2014). doi: 10.1016/j.trc.2014.01.005.
20. Wang, Y. *et al.* "Error Assessment for Emerging Traffic Data Collection Devices," (June), p. 106 (2014). Available at: http://www.wsdot.wa.gov/research/reports/fullreports/810.1.pdf.
21. Williams, B. M. and Hoel, L. A. "Modeling and forecasting vehicular traffic flow as a seasonal ARIMA process: Theoretical basis and empirical results," *Journal of Transportation Engineering*, 129(6), pp. 664–672 (2003). doi: 10.1061/(ASCE)0733-947X(2003)129:6(664).
22. Yang, M., Liu, Y. and You, Z. "The reliability of travel time forecasting," *IEEE Transactions on Intelligent Transportation Systems*, 11(1), pp. 162–171 (2010). doi: 10.1109/TITS.2009.2037136.
23. Yin, H. *et al.* "Urban traffic flow prediction using a fuzzy-neural approach," *Transportation Research Part C: Emerging Technologies*, 10(2), pp. 85–98 (2002). doi: https://doi.org/10.1016/S0968-090X(01)00004-3.